\documentclass[useAMS,usenatbib,usegraphicx]{mn2e}

\usepackage{times}
\usepackage{epsfig}
\usepackage{epstopdf}
\usepackage{graphicx}
\usepackage{color}
\usepackage{txfonts}

\def\kms{\relax \ifmmode {\,\rm km\,s}^{-1}\else \,km\,s$^{-1}$\fi}
\def\ha{\relax \ifmmode {\rm H}\alpha\else H$\alpha$\fi}
\def\hb{\relax \ifmmode {\rm H}\beta\else H$\beta$\fi}
\def\hi{\relax \ifmmode {\rm H\,{\sc i}}\else H\,{\sc i}\fi}
\def\hii{\relax \ifmmode {\rm H\,{\sc ii}}\else H\,{\sc ii}\fi}
\def\h2{\relax \ifmmode {\rm H}_2\else H$_2$\fi}
\def\degr{\hbox{$^\circ$}}
\def\arcmin{\hbox{$^\prime$}}

\newcommand{\mum}{$\umu$m}

\newcommand{\msol}{\mbox{\,$M_\odot$}}        
\def\hmsd#1h#2m#3.#4s{                  
                      \relax
                      \ifmmode #1^{\rm h}\,#2^{\rm m}\,#3\fs#4
                      \else \hbox{$#1^{\rm h}\,#2^{\rm m}\,#3\fs#4$}
                      \fi
                     }
\def\dmsd#1d#2m#3.#4s{                  
                      \relax
                      #1\degr\,#2\arcmin\,#3\farcs#4
                     }

\title[Interactions and star formation]{Interacting galaxies in the nearby Universe: only moderate increase of star formation}

   \author[Johan H. Knapen et al.]{
   	   Johan H. Knapen$^{1,2}$,
	   Mauricio Cisternas$^{1,2}$,
	   Miguel Querejeta$^{3}$\\
$^{1}$Instituto de Astrof\'\i sica de Canarias, E-38200 La Laguna, Tenerife, Spain\\
$^{2}$Departamento de Astrof\'\i sica, Universidad de La Laguna, E-38205 La
Laguna, Tenerife, Spain\\
$^{3}$Max Planck Institute for Astronomy, K\"onigstuhl 17, 69117, Heidelberg, Germany}

\begin{document}

\date{Accepted 2015 September 14.  Received 2015 August 7; in original form 2015 February 13}

\pagerange{\pageref{firstpage}--\pageref{lastpage}} \pubyear{2014}

\maketitle

\label{firstpage}

\begin{abstract}

We investigate the influence of interactions on the star formation by studying a sample of almost 1500 of the nearest galaxies, all within a distance of $\sim45\,$Mpc. We define the massive star formation rate (SFR), as measured from far-IR emission, and the specific star formation rate (SSFR), which is the former quantity normalised by the stellar mass of the galaxy, and explore their distribution with morphological type and with stellar mass. We then calculate the relative enhancement of these quantities for each galaxy by normalising them by the median SFR and SSFR values of individual control populations of similar non-interacting galaxies. We find that both SFR and SSFR are enhanced in interacting galaxies, and more so as the degree of interaction is higher. The increase is, however, moderate, reaching a maximum of a factor of 1.9 for the highest degree of interaction (mergers). The SFR and SSFR are enhanced statistically in the population, but in many individual interacting galaxies they are not enhanced at all. We discuss how those galaxies with the largest SFR and/or SSFR enhancement can be defined as starbursts. This study is based on a representative sample of nearby galaxies, including many low-mass and dwarf/irregular galaxies, and we argue that it should be used to place constraints on studies based on samples of galaxies at larger distances, beyond the local Universe.

\end{abstract}

\begin{keywords}
{Galaxies: interactions --- Galaxies: starburst  --- Galaxies: spiral}
\end{keywords}


\section{Introduction}

Galaxy-galaxy interactions and mergers are fundamental in our current thinking of how the Universe has evolved since its very earliest stages. In the accepted cosmological framework mergers between dark matter haloes and/or luminous galaxies increase the mass of the haloes and galaxies, and shape the galaxies from more irregular, clump-like structures, to the generally smooth and/or disk-like bodies that we observe at the current epoch. One assumption that is often made is that interactions and in particular mergers stimulate enhanced star formation, leading to star formation rates (SFRs) that are temporarily increased, often by very large amounts.

That this happens is a certainty. Some of the objects with the most extreme SFRs are selected by their large infrared emission and are known as (Ultra-)Luminous InfraRed Galaxies, or (U)LIRGs, and it is well known that in particular the more powerful ULIRGs are almost without exception interacting or merging galaxies (e.g., Joseph \& Wright 1985; review by Sanders \& Mirabel 1996). At higher redshifts this kind of objects may well be more common, but at redshift zero ULIRGs are exceedingly rare (what is considered to be the closest ULIRG, Arp~220, is at $\sim70$\,Mpc). Numerical simulations can and do often produce much enhanced SFRs as a result of mergers (e.g., Barnes \& Hernquist 1991; Mihos \& Hernquist 1994; Bournaud et al. 2011).

It is vital to establish, however, whether a galaxy-galaxy interaction or merger is always accompanied by a significant increase in the star formation activity, in other words, whether they lead to statistical enhancements of the SFR. Various observational results indeed show this, but find a rather limited increase in the SFR, of factors of a few (e.g., Larson \& Tinsley 1978; Bergvall et al. 2003; Robaina et al. 2009; Knapen \& James 2009, hereinafter KJ09; Ellison et al. 2013; Barrera-Ballesteros et al. 2015; Brassington et al. 2015), or, in a similar fashion, a limited decrease in gas depletion time (e.g., Saintonge et al. 2012). Such limited SFR enhancement is confirmed by numerical modelling (e.g., Kapferer et al. 2005; Di Matteo et al. 2007, 2008; Moreno et al. 2015). The SFR enhancement depends on parameters such as the relative masses of the interacting galaxies, the separation between galaxies in close pairs, and the galaxy environment (e.g., Ellison et al. 2008; 2010).

One outstanding issue is sample selection and resulting biases in the measured parameters. A sample can be too small to include statistically significant numbers of merging or interactions (e.g., KJ09), or studies can be based on samples which are large but which may miss the population of faint dwarfs and irregular galaxies observed near our own Milky Way (e.g., most studies of samples of galaxies at moderate or higher redshift), or a combination of these and many other factors. These lower-mass galaxies are important, as are minor mergers (e.g., Kaviraj 2014), because they contribute significantly to the evolution of the overall galaxy population (e.g., Lelli, Verheijen \& Fraternali 2014; Stierwalt et al. 2015). 

In this paper, we study the statistical SFR and specific SFR (SSFR; SFR normalised by stellar mass) of a sample of some 1500 galaxies in the very local Universe, within a distance of $\sim$45\,Mpc, by combining data obtained as part of the {\it Spitzer} Survey of Stellar Structure in Galaxies (S$^4$G, Sheth et al. 2010) with knowledge of the interaction properties of these galaxies, from Knapen et al. (2014, hereinafter Paper~I). This sample uniquely covers galaxies down to stellar masses of around $10^8\msol$ and of all morphological types, from elliptical to irregular. For all these galaxies we have reliable SFR and stellar mass determinations, and have determined their interaction class from detailed optical imaging. 


\section{Sample selection and data}

\begin{figure}
\includegraphics[width=0.45\textwidth]{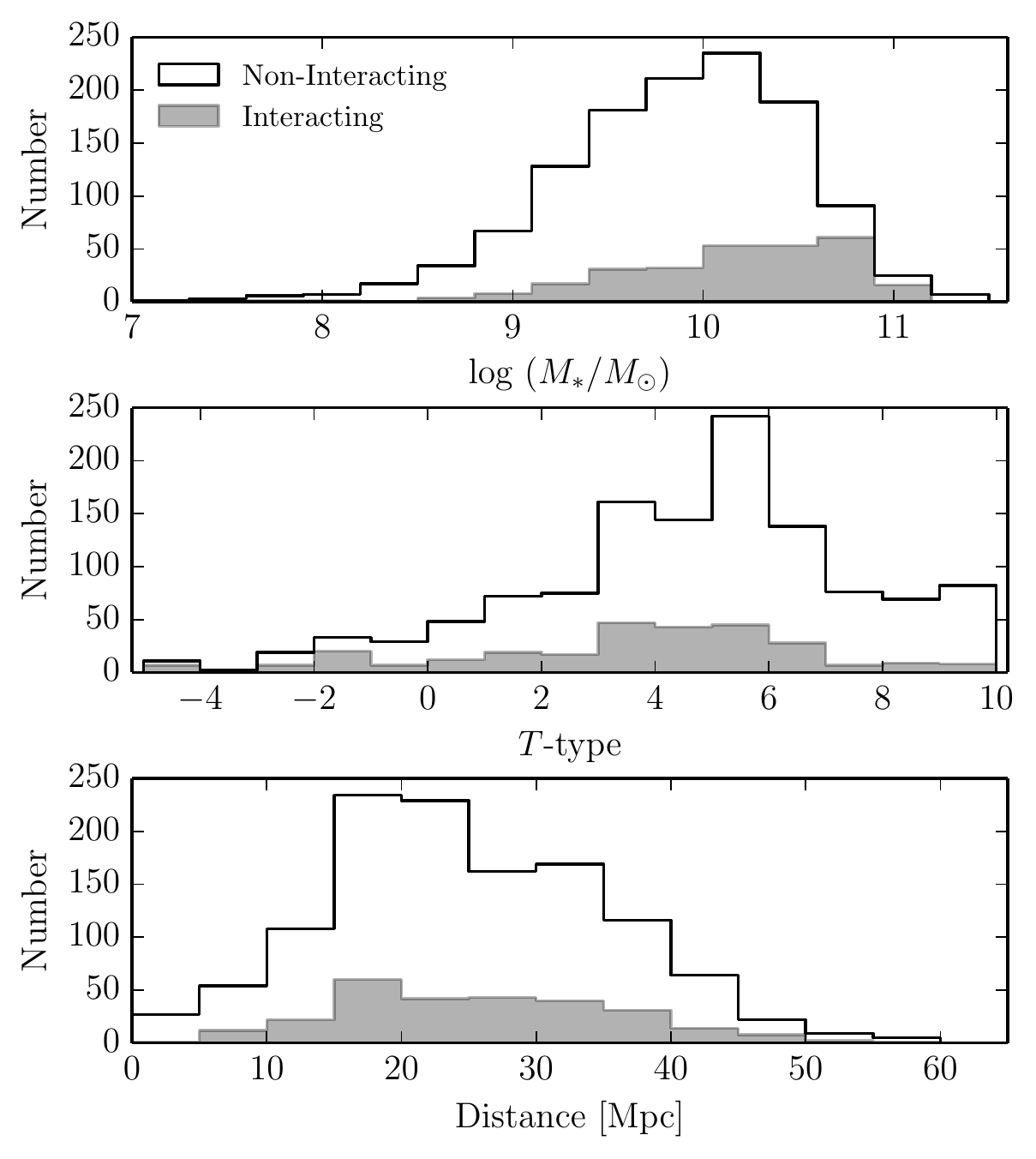}
\caption{Histograms showing the distribution of the sample galaxies in terms of stellar mass (top), morphological {\it T}-type (middle) and distance (lower panel). Stellar mass is given in units of Solar mass, type follows the convention established in the RC3, where $T=0$ is an S0, $T=1$ S0a, $T=2$ Sa, etc., and distance is given in units of Mpc. The non-interacting and interacting galaxies in the sample are plotted separately.}
\label{sample}
\end{figure}

\begin{figure*}
\includegraphics[width=0.8\textwidth]{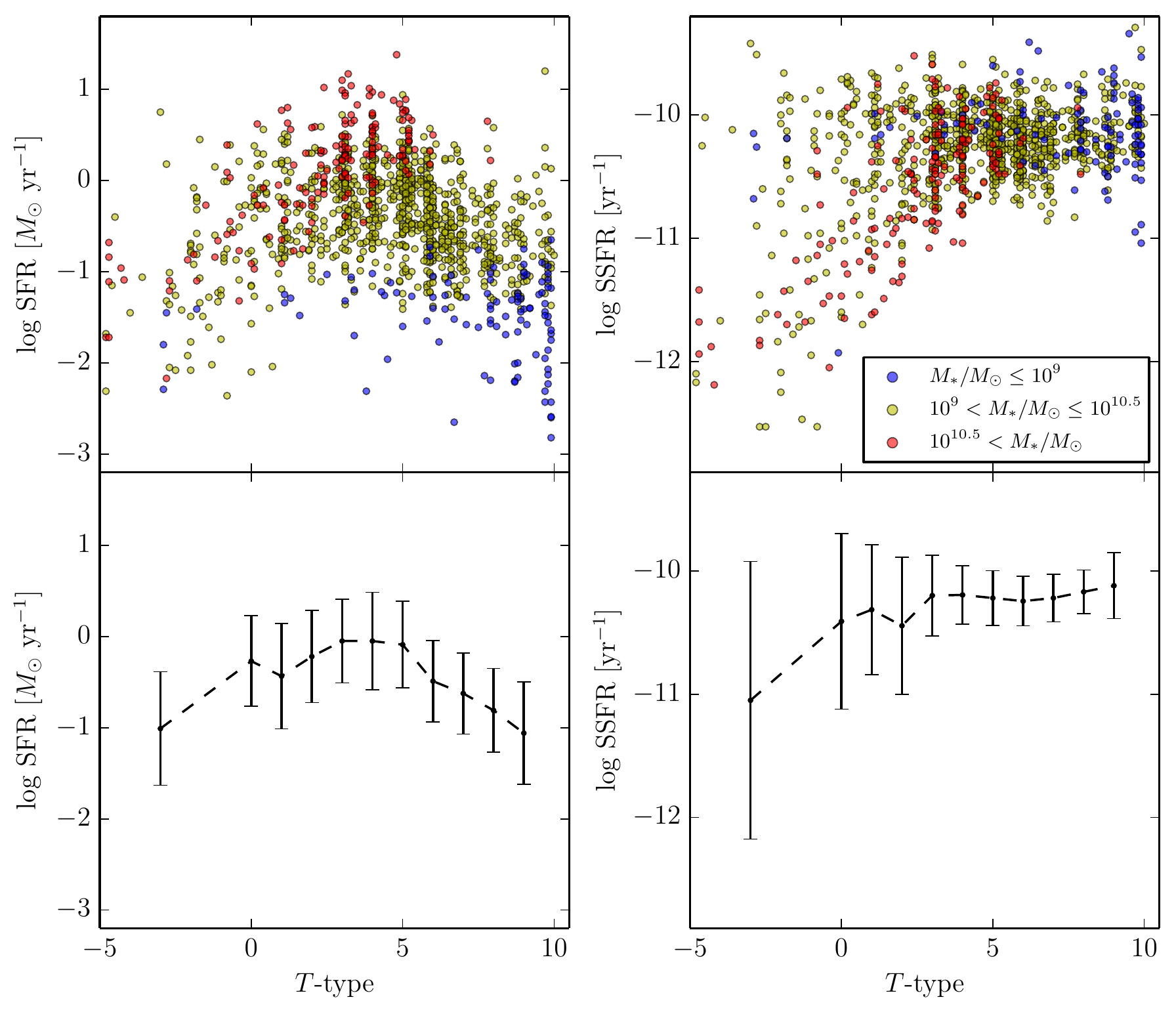}
\caption{Log of the SFR (left panels) and the SSFR (right panels) as a function of morphological type. Top panels show the individual galaxies, colour-coded by galaxy stellar mass, while the lower panels show the median values, with the $1\,\sigma$ spread indicated by the error bars. The dashed line connects the points.}
\label{sfrtype}
\end{figure*}

We take advantage of well-resolved imaging and detailed measurements of a number of key parameters which we have derived in related papers for a substantial number of nearby galaxies. In particular, we use for some 1500 galaxies from the S$^4$G sample (Sheth et al. 2010) information on: 1) whether they have close companions, or are interacting (from Paper~I), 2) the stellar mass, determined from dust emission-corrected 3.6\,\mum\ imaging, and 3) the SFR, determined from {\it IRAS} ({\it InfraRed Astronomical Satellite}) photometry at 60 and 100\,\mum. Both stellar mass and SFR are from Querejeta et al. (2014), and are combined to give measures of the SSFR. We now give the basics on how the sample and these parameters  were derived, but in any case refer to the original papers for a more complete description. 

{\it Input sample}: The original S$^4$G sample consists of 2352 nearby, bright and large galaxies ($d \lesssim45$\,Mpc, $m_B < 15.5, D_{25} > 1$\,arcmin, as obtained from HyperLEDA, and $|b|>30$\,deg). These were selected using radial velocity information in the literature, which led to a deficiency of early-type, gas-poor galaxies. \footnote{A newly approved extension to the S$^4$G survey will increase the size of the  sample to a total of 2829 nearby galaxies, covering the whole sky, and selected using the above limits in volume, magnitude, size, and Galactic latitude. Images at 3.6 and 4.5\,\mum\ obtained with {\it Spitzer} are already publicly available for the original S$^4$G sample, and with Paper~I we released optical imaging for 1768 galaxies in the extended S$^4$G sample. Stellar masses and {\it IRAS} SFRs are, however, not yet available for the galaxies in the extension to the original S$^4$G sample.}

{\it Companions and interactions}: In Paper~I, we studied the presence of companion galaxies near the 2829 galaxies in the extended S$^4$G survey. As a first step, we used the NASA-IPAC Extragalactic Database (NED) to find all close companion galaxies, following the criteria of KJ09, namely those which (1) are within a radius (measured from the centre of the galaxy) of five times the diameter of the sample galaxy, or $r_{\rm comp} < 5 \times D_{25}$, with $D_{25}$ from de Vaucouleurs et al. (1991; RC3), {\it and} (2) have a recession velocity within a range of $\pm200$\,km\,s$^{-1}$ of the galaxy under consideration, {\it and} (3) are not more than 3 mag fainter (using magnitudes in NED). These criteria ensure that galaxies form physical pairs, and are massive enough to exert a noticeable gravitational pull on each other.

We then visually inspected all galaxies classified as having a close companion in order to assign them an interaction class. Here we used three classes, namely A. mergers, B. highly distorted galaxies, and C. galaxies with minor distortions, with the sequence C, B, and A indicating increasing levels of interaction. As discussed in considerable detail in Paper~I, the criteria employed lead to a classification which is comparable to other approaches used in the literature. Galaxies with close companions are in the 0 category, but are hardly considered at all in what follows.

All non-interacting galaxies for which we could not identify a close companion are placed in the control sample category, which we call N. 

{\it Stellar masses}: We obtain stellar masses for our sample galaxies from Querejeta et al. (2014). They derive these masses from S$^4$G 3.6\,\mum\ images after correcting them for dust emission which can contribute to a significant degree to the 3.6\,\mum\ flux, using the Independent Component Analysis presented by Meidt et al. (2012). The resulting images, which trace primarily old stars, can then be transferred into mass by integrating them within the 25.5\,mag/arcsec$^2$ isophote, and using a mass-to-light ratio ($M/L$) of 0.6. This value has been proposed by Meidt et al. (2014) using an empirical calibration based on stars from the GLIMPSE survey, and has been ratified independently by R\"{o}ck et al. (2015) on the basis of new stellar population synthesis modelling extending to the range of 2.5-5\,\mum\ in wavelength by using empirical stellar spectra. R\"{o}ck et al. conclude that a mass-to-light ratio of 0.6 is a decent average value to use, although it does depend on age, and in particular on the shape and slope of the IMF (though hardly at all on metallicity).  Norris et al. (2014) provide a third independent confirmation, from $WISE$ near-IR colours of a diverse sample of dust-free stellar systems used to validate the efficacy of a range of stellar population synthesis models, that indeed a value of $M/L=0.6$ is a reasonable estimate, to a precision of $\sim 0.1$\, dex. 

{\it SFRs and SSFRs}: SFRs were also taken from Querejeta et al. (2014), who derived these from the weighted averages of the {\it IRAS} 60 and 100\,\mum\ fluxes as reported in the NED, converting them using the recipe from Larsen \& Richtler (2000). Dividing the SFR by the stellar mass of a galaxy yields the SSFR. 

Using {\it IRAS} fluxes to estimate SFRs is established practice (see, e.g., Buat \& Xu 1996; Charlot et al. 2002; review by Calzetti 2013), and leads to acceptable values for large samples of galaxies. Alternatives might be better but would in our case have led to severe disadvantages. For instance, using combinations of some or all of {\it IRAS} IR, {\it Spitzer} 8 and 24\,\mum, or {\it GALEX} FUV fluxes (see, e.g., Calzetti 2013) leads to arguably better SFR estimates, but would reduce our sample size significantly, so much so, in fact, as to make it useless for our statistical purposes. For the same reason, the use of H$\alpha$ images to estimate SFRs is excluded, as such images are only available for around 25\% of the S$^4$G sample.

We compared the results of this paper as presented (based on {\it IRAS}-based SFRs) with those derived using {\it GALEX}-based SFRs (not corrected for dust extinction), which are available for most of the S$^4$G galaxies. As might reasonably be expected, we find that the {\it IRAS}-based median SFRs and SSFRs are lower for low-mass galaxies and higher for high-mass ones, relative to the {\it GALEX}-based values. This is partly because of selection effects, with {\it IRAS} detection rates higher for high-mass and lower for low-mass galaxies relative to {\it GALEX} detection rates. We repeated the complete analysis as presented in this paper with UV- instead of IR-based SFRs, and found that qualitatively they results are identical, while quantitatively they are slightly different. As our overall results and conclusions, as presented below, do not change significantly when using {\it GALEX} SFRs rather than {\it IRAS} ones, we chose to use {\it IRAS} SFRs mainly because they are less affected by dust extinction than UV SFRs, with dust extinction effects being the most worrisome when studying interactions and increased SFRs.

An additional potential worry is the rather low spatial resolution of {\it IRAS}.  It is hard to give one number for the {\it IRAS} ``beamsize", but we estimate that around half of our Class A interacting galaxies are so close together that they will not have been observed separately by {\it IRAS}. Our Class B and C galaxies are considerably less close, and only around 10\% of those would not be separated. For non-separated galaxy pairs, the SFR derived will be the sum of the SFRs of the individual galaxies. In principle, this could result in artificially increased values of the SFR excess (compared to a control sample of non-interacting galaxies; Fig.~\ref{enhan} and Sect.~3.3). However, this artificial increase would only occur for our Class A, not for Class B or C. And as Class A has the same median (S)SFR excess as Class B (Fig.~\ref{enhan}, lower panels), we argue that the large ``beam" of {\it IRAS} does not affect our overall results. The physical background behind this is probably that in many cases only one of two interacting galaxies is actively forming stars (see also KJ09), while in a few other galaxies the merger is so advanced that two original galaxies cannot easily be distinguished at all. We conclude that the low {\it IRAS} spatial resolution has no significant effect on our results, a conclusion also supported by the fact (reported above) that we find the same results when using {\it GALEX} imaging at much better spatial resolution to derive the SFRs.

We do not expect contributions from AGN to make a significant statistical impact, because few of our sample galaxies have AGN with powerful IR emission (Cisternas et al. 2013).

{\it Final sample}: The final sample used for our analysis is smaller than the input S$^4$G sample primarily because {\it IRAS} fluxes are not available for all galaxies, and because the stellar masses and SFRs from Querejeta et al. (2014) are for the time being only available for a subset of the original S$^4$G sample. Our final sample consists of 1478 nearby galaxies, at distances smaller than $\sim45\,$Mpc (the median distance across the sample is 23.7\,Mpc, whereas only 3.5\% of galaxies have distances larger than 45\,Mpc, and 1.5\% are at $D>50$\,Mpc). The distribution of the non-interacting and interacting galaxies in our sample as a function of stellar mass, morphological {\it T}-type, and distance is shown in Figure~\ref{sample}. The distribution of both sub-samples is very similar, and covers a broad spread in galaxy properties. We highlight in particular the substantial number of sample galaxies with relatively low masses, and at late types. The morphological type has been taken from the RC3, but the classifications change remarkable little when done on the basis of 3.6\,$\mu$m images (Buta et al. 2010, 2015). 

We have checked whether the reduction in number of galaxies from the input to the final sample might affect the statistics of interacting galaxies, and found that this is not the case. The fractions of galaxies in the various interaction classes are the same within the uncertainties for those galaxies in our input sample that do and those that do not have {\it IRAS} fluxes and stellar masses in Querejeta et al. (2014). The only difference that is possibly just significant is a lower number of galaxies with companions (but not in our interaction classes A, B, or C) among the 792 S$^4$G galaxies which are not in the final sample considered in the current paper. This is Class 0 in Paper~I, which is not considered in much detail in the current paper. 

We thus conclude that the sample analysed here properly represents the characteristics of the galaxy population in the nearby Universe.


\section{Results}

\subsection{Overall statistics}

Of the 1478 galaxies of the reduced sample for which we have all the necessary information, 16 are of type A, 39 of B, 84 C, 138 of type 0 (with close companion but not interacting), and the remaining 1201 galaxies are type N which do not have a close companion. The latter category will in subsequent analyses be used as a reservoir from which to construct control samples.

To get an idea of the distribution of galaxies within our sample, we first show the spread of SFR and SSFR with morphological type, as obtained from the RC3 (shown in Figure~\ref{sfrtype}, top panels, and as a run of the median values for each type in Figure~\ref{sfrtype}, lower panels). The figures show, first of all, that our sample consists of galaxies of all morphological types, including ellipticals and irregulars. Our selection criteria, as discussed in the previous Section, may have introduced biases against certain types of galaxies, in particular quiescent early-types, but this bias is not so large as to exclude certain galaxy classes altogether. 

The figures also confirm the well-known fact that the SFR is highest in mid-type galaxies (gas-rich spirals) and lower in galaxies of the earliest and latest types. The spread in SFR for individual galaxies is rather large, however.

The distribution of SSFRs, unlike that of SFRs, is flat at later morphological types. This is because the decline in SFR there is offset by the lower stellar masses in these galaxies, with the result, again well-known from previous work, that the SSFR is lower for early-type galaxies, but rather constant for all disk galaxies (of type later than say $T=1$).

\begin{figure*}
\includegraphics[width=0.8\textwidth]{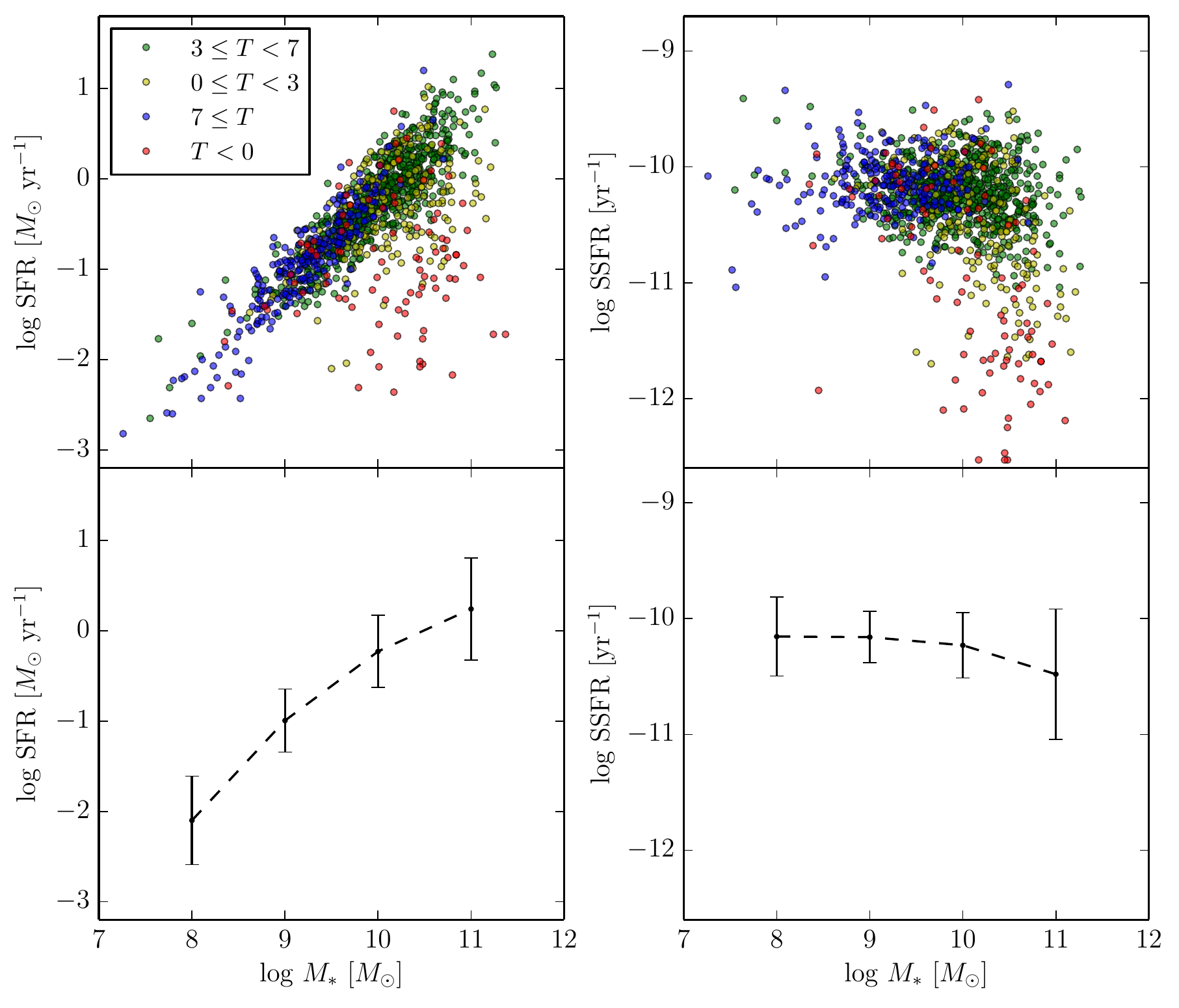}
\caption{As Figure~\ref{sfrtype}, but now for the stellar mass, and colour-coded by morphological type.}
\label{sfrmass}
\end{figure*}

Figure~\ref{sfrtype} thus confirms the overall characteristics of our sample, and shows that, at least statistically, the approach taken to determine key parameters like SFR and SSFR is valid. This is further confirmed when we plot the SFR and SSFR against the stellar mass of the sample galaxies, as in Figure~\ref{sfrmass}. Again these are well-known plots, of which the SFR vs mass one is popularly known as the 'star formation main sequence'. And again we see that our sample reproduces this nicely: the SFR rises with stellar mass of the galaxy,  whereas the SSFR, which is of course SFR normalised by mass, is rather flat over most of the mass range, but with a noticeable tail of low-SSFR galaxies at high masses: large galaxies with low SFRs, typically of early type (see colour-coding in Figure~\ref{sfrmass}). As in the case of SFR vs morphological type, the range in values for a particular stellar mass is rather large (typical median absolute deviations

The stellar mass of our sample galaxies ranges from roughly $10^8$ to $10^{11}$\,\msol, with very few galaxies at lower or higher masses. Comparing this mass range with other studies based on many more galaxies, for instance using galaxies from the Sloan Digital Sky Survey (SDSS), we observe that SDSS-based studies typically have fewer galaxies at lower masses, and more at higher masses than us (as examples, compare our Figure~\ref{sample} with Figure~1 of Ellison et al. 2010, or the stellar mass distribution of our sample as shown in Figure~\ref{sfrmass} with that shown in Figure~1 of Luo et al. 2014). This can be explained by the fact that our sample galaxies are closer, and that most SDSS or higher-redshift studies will miss smaller and catch more large galaxies as the distance and redshift of the sample increases. This is in fact a major strength of our sample.

Another feature that is obvious when comparing our Figure~\ref{sfrmass} with typical SDSS-based star formation main sequence plots in the literature is that our sample preferentially populates the so-called blue sequence, and that it is poor in 'red and dead' galaxies. This is most probably due to our selection criteria in combination with the point raised in the previous paragraph, that high-mass galaxies are rarer in the local Universe than they are at higher redshifts. In any case, it does in no way devalue our study as typical SDSS-based studies of SFR properties concentrate on the blue sequence.

\subsection{Control samples}

In order to investigate whether, and by how much, interactions between galaxies change the SFR and SSFR, we must first define a control sample with which the interacting galaxies can be compared. Taking advantage of the fact that we have parameters, including SFR and SSFR, for 1201 non-interacting galaxies without close companions, we have chosen an approach where the control values of SFR and SSFR of each interacting galaxy are the median values of the SFRs and SSFRs of all non-interacting galaxies which resemble the interacting galaxy, in the sense that they are similar in morphological type and stellar mass (type within $\pm1$ numerical class, stellar mass within $\pm0.2$ in $\log(M/M_{\odot})$). For each of the galaxies in our companion or interaction classes 0, A, B, or C, we then divide the SFR and SSFR by the median value of the SFR and SSFR of their respective control group of galaxies. The size of this control group varies from 7 to 265, with most interacting galaxies having control groups of near 100 (mean: 113, median: 105). 

\subsection{SFR and SSFR enhancement due to interactions}

\begin{figure*}
\includegraphics[width=0.8\textwidth]{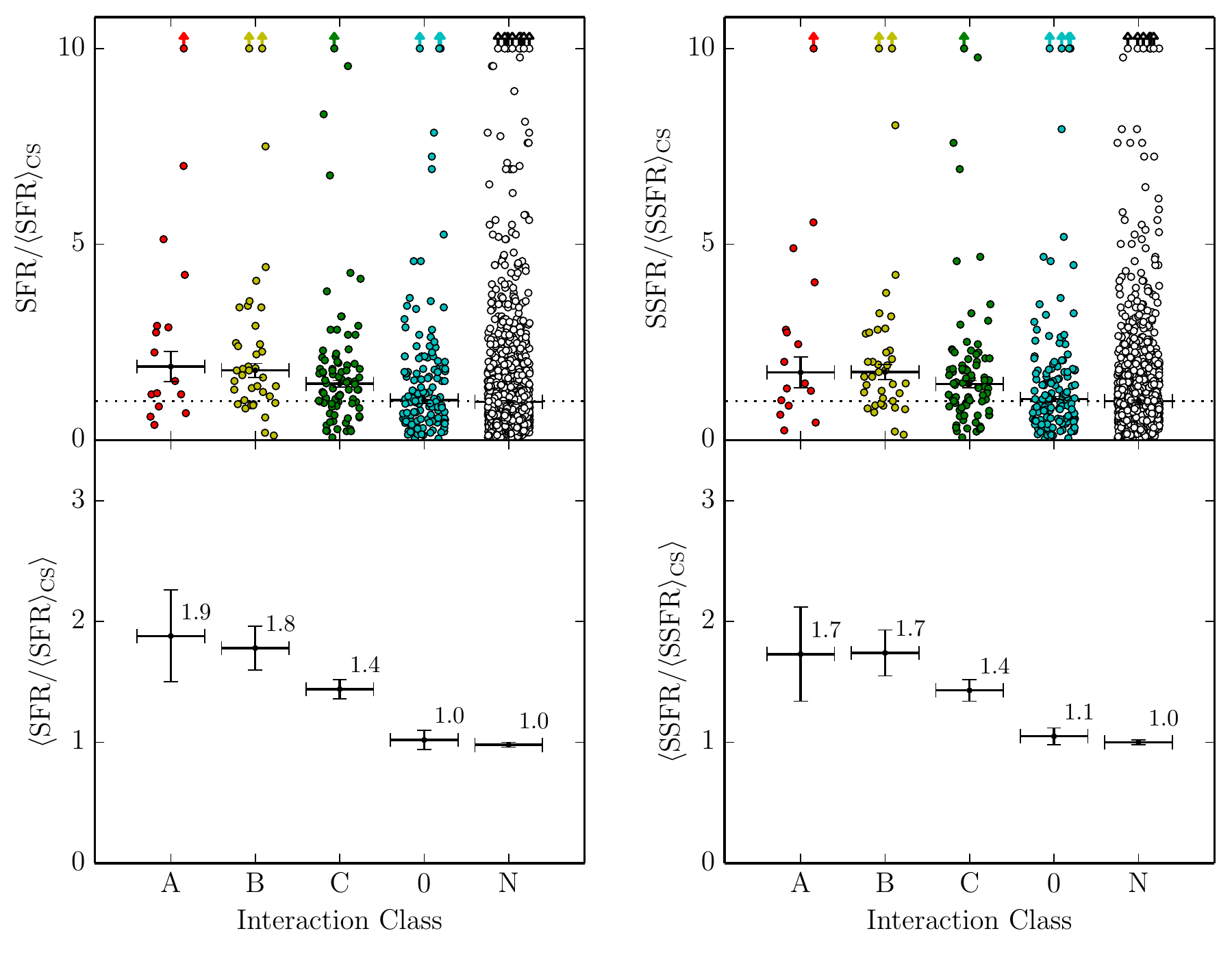}
\caption{The SFR (left panels) and SSFR (right panels) enhancement, defined as the (S)SFR of a galaxy normalised by the median (S)SFR for its control population (see Sect.~3.2), separated by interaction class (where A is the most extreme class, of mergers, and N contains those galaxies which are neither interacting nor have a close companion). Values larger than 10 are indicated by lower limits, and are listed individually in Table~\ref{extremes}. Median values per class are indicated with their $1\,\sigma$ uncertainty in both top and lower panels, but are amplified in the lower panels where the values are indicated next to the data points.}
\label{enhan}
\end{figure*}

\begin{table*}
\begin{center}
\begin{tabular}{llcccccc}
\hline
\noalign{\smallskip} 
Class & Galaxy & $T$-type & log $M_{\ast}$ & log SFR & log SSFR &SFR/$\langle{\rm SFR}\rangle_{\rm CS}$ & SSFR/$\langle{\rm SSFR}\rangle_{\rm CS}$\\
 & & & [$M_{\odot}$] & [$M_{\odot}$ yr$^{-1}$] & [yr$^{-1}$] & & \\
\hline
\noalign{\smallskip}
A & NGC 0520 &  0.8 & 10.66 &  1.08 & $-$9.58 & 22.1 & 22.1 \\
B & NGC 2798 &  1.1 & 10.24 &  0.82 & $-$9.42 & 10.7 & 12.3 \\
 & NGC 5018 & $-$4.6 & 11.14 & $-$0.13 & $-$11.27 & 18.8 & 20.2 \\
C & NGC 1482 & $-$0.9 & 10.20 &  0.79 & $-$9.41 & 17.8 & 15.7 \\
0 & NGC 3065 & $-$2.0 & 10.46 & $-$0.08 & $-$10.54 & 15.5 & 14.8 \\
  & NGC 4355 &  1.0 &  9.82 &  1.04 & $-$8.78 & 21.9 & 20.0 \\
  & NGC 5173 & $-$4.7 & 10.07 & $-$0.59 & $-$10.66 &  (7.2) & 10.2 \\
 & NGC 7465 & $-$1.9 &  9.99 &  0.25 & $-$9.74 & 12.5 & 13.2 \\
N  & ESO 402-030 & $-$1.2 & 10.02 &  0.15 & $-$9.87 &  (9.8) & 12.3 \\
  & NGC 1222 & $-$3.0 & 10.17 &  0.75 & $-$9.42 & 114.8 & 164.7 \\
  & NGC 1808 &  1.2 & 10.55 &  0.80 & $-$9.75 & 11.0 &  (7.6) \\
  & NGC 2764 & $-$1.7 & 10.31 &  0.45 & $-$9.86 & 28.2 & 36.7 \\
  & NGC 3094 &  1.1 & 10.50 &  0.90 & $-$9.60 & 13.8 & 10.0 \\
  & NGC 3928 & $-$4.5 &  9.62 & $-$0.40 & $-$10.02 & 81.3 & 120.2 \\
  & NGC 4194 &  9.7 & 10.49 &  1.20 & $-$9.29 & 11.2 &  (3.8) \\
  & NGC 4385 & $-$0.7 & 10.19 &  0.39 & $-$9.80 &  (6.9) & 12.0 \\
  & NGC 5078 &  1.0 & 11.15 &  0.77 & $-$10.38 & 10.7 &  (5.3) \\
  & NGC 6014 & $-$1.8 & 10.17 & $-$0.19 & $-$10.36 & 10.5 & 16.6 \\
\noalign{\smallskip}
\hline
\end{tabular}
\end{center}
   \caption[]{Extreme values of SFR and SSFR, listing all galaxies for which these are enhanced by a factor $\geq10$ compared to their control population. Listed are interaction class, galaxy identification, morphological type, stellar mass, SFR and SSFR, and the SFR and SSFR enhancements. Values $<10$ in the any of last two columns are in parentheses.}
\label{extremes}
\end{table*}

The SFR and SSFR enhancements in the various classes of galaxies, which are their SFRs and SSFRs normalised to the control values (see Sect. 3.2) are plotted for all individual galaxies in the top panels of Figure~\ref{enhan}, and as median values for each interaction class in the lower panels of Figure~\ref{enhan}. Various inferences can be drawn from these figures:

\begin{itemize}

\item There is a huge spread in enhancement of both SFR and SSFR, in all classes, and including in the control class (N).

\item The maximum SFR and SSFR values are very large indeed compared to the median values. Values higher than 10 have been indicated in the figures with upward-pointing arrows, and are explicitly listed in Table~\ref{extremes}. This Table contains a number of well-known extreme galaxies, such as NGC~1222 (starburst), but the fact that Table~\ref{extremes} only contains 18 entries indicates that such galaxies are very rare.

\item The median values for the `N' sample are unity, confirming that the approach we have taken to calculate the (S)SFRs normalised by control groups is valid.

\item The median SFR and SSFR enhancements are higher than unity in interacting galaxies, but only moderately so. Even for the most extreme of our categories (A, mergers) the median enhancement is a mere factor of 1.9. The median SSFR is enhanced by very similar factors.

\item Both the median SFR and SSFR enhancements increase with interaction class. The highest enhancements are found in the closest interaction stages (A, B, C, in that order). The fact that the peaks in enhancement occur near coalescence is perhaps not unexpected, but because major mergers are highly chaotic events it is hard to observe this effect in individual cases, or in small samples. 

\end{itemize}

\subsection{Starburst definition and fractions}

\begin{figure*}
\includegraphics[width=0.8\textwidth]{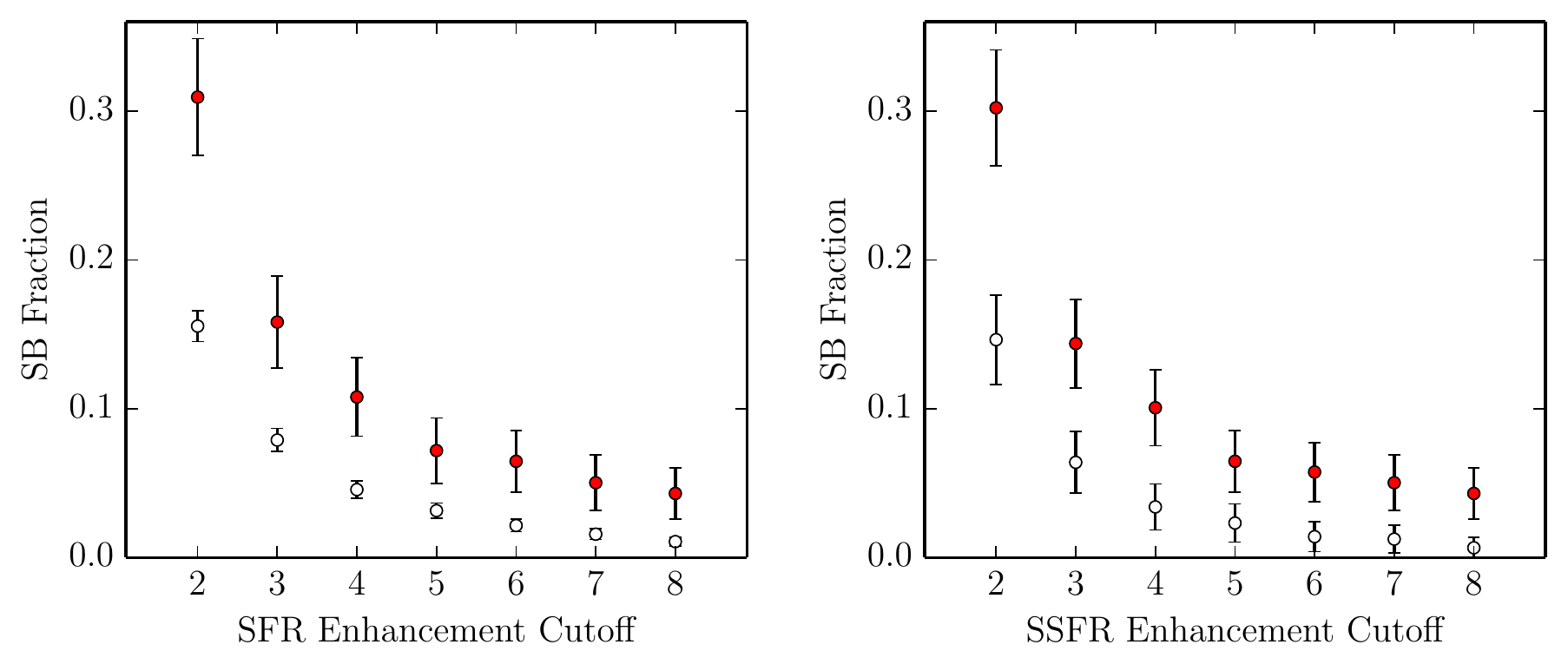}
\caption{Variation of the fraction of galaxies classified as starburst, as a function of the definition used for starburst, in terms of cutoff level in SFR (left panel) and SSFR enhancement (right). Filled red dots indicate the starburst fraction among those galaxies which are interacting (our classes A, B, and C), open dots that among all galaxies. In this paper, we suggest that using values of 5 in SFR and 4 in SSFR are reasonable to define starbursts---Table~\ref{extremes} lists those few galaxies selected when a more extreme value of 10 is used.}
\label{SB-fraction}
\end{figure*}

Figure~\ref{SB-fraction} shows what fraction of galaxies (filled red dots, of the 139 interacting galaxies, groups A, B and C; and open dots, of the total sample of 1478 galaxies) is classified as starburst using different values of the SFR (left panel) and SSFR (right) enhancements; obviously these fractions go down with more stringent criteria. These cutoff values correspond to imaginary horizontal lines in Figure~\ref{enhan}, where the dotted line which is plotted there indicates an enhancement of unity. The combination of these two figures illustrates one of the tradeoffs in deciding what a reasonable definition for  starburst is: the stricter the better as more extreme objects will be selected, but on the other hand stricter criteria will restrict the number of galaxies classified as starbursts, to so few as to be useless as a class for the most restrictive criteria.

We argue that a reasonable value to use here is an enhancement of 5 in terms of SFR, and of 4 in terms of SSFR (lower in the latter case simply because SSFR enhancement values are generally lower, see lower panels of Figure~\ref{enhan}). This definition yields some 50 `starburst' galaxies in total, of which some 15 are in interacting galaxies---numbers which allow statistical analysis while at the same time ensuring that the classification of starburst remains exclusive. Similar values are used in the literature, for instance by Rodighiero et al. (2011) or Luo et al. (2014), while the consequences of using other definitions for a starburst have been discussed by us in other papers, in particular KJ09 and Knapen \& Cisternas (2015, hereinafter Paper~III). Note that in Table~\ref{extremes} we only list those galaxies with enhancement values higher than 10. This represents a much more restrictive and extreme starburst definition, and the main reason to include them in the table is in fact not more profound than that these are the galaxies that are plotted as lower limits in Figure~\ref{enhan}.


\section{Discussion}

\subsection{Interactions enhance the star formation}

The first main result is that statistically, interactions do enhance the SFR of galaxies. This is already a well-established result, obtained from both anecdotal or statistical studies. The general statement on enhancing the SFR must, however, be moderated by two related results, both of which can be seen clearly in Figure~\ref{enhan}, namely (1) that there is a huge range in the fractions by which the SFR is enhanced in individual galaxies, with tremendous overlaps between the different sub-samples of interacting or non-interacting galaxies, and (2) that the median value of the SFR enhancement related to interactions is relatively modest, only a factor 1.9 in the case of our most extreme interaction class, those of the mergers. 

The huge range in the factor by which the SFR is enhanced in individual galaxies is detailed in Table~\ref{extremes}, where we list those galaxies where the SFR is more than 10 times larger than in their control population. The origin of these high values is explored in some more detail below, in Sect.~4.3, but the point we note here is that the two highest values occur in the control sample (in NGC~1222 and NGC~3928), not among the interacting galaxies. So extreme SFRs can and do occur in the absence of interactions, as well as in interacting galaxies. In the latter, the high SFR may well be caused by the interaction, as shown in many numerical modelling studies (see references in Sect.~1). It is also well known that among starburst galaxies, and in particular among the most extreme ones, interactions are more, or very, common. We find and discuss this on the basis of our data in Paper~III), and for extreme starbursts there is a large body of supporting work in the literature, in particular related to (U)LIRGs. 

The important result here is thus the statistical enhancement of the SFR by interactions, in median values across samples, rather than enhancements in individual galaxies. But we confirm earlier works and find that this enhancement is modest. For instance, similar enhancements, by factors of a few, were found from numerical modelling by Di Matteo et al. (2007, 2008) and from observations by KJ09. In the latter study, we did not have a large enough sample to consider mergers as a separate class, and instead reported a SFR enhancement of a factor 2.5 among galaxies with close companions. Here, our close companion subsample (class 0) does not show enhanced SFRs, which is somewhat surprising. We postulate that this is due to sample selection, in the sense that the uncertainties on the KJ09 result were larger due to the smaller sample size, but also in the sense that in the present work we have been able to separate those galaxies which are interacting (classes A, B, C) from those which do not, even though they have a close companion (class 0). 

There has been some debate in the literature regarding the physical mechanisms, scales, and numerical implementation of the SFR enhancements in merging and interaction galaxies. Di Matteo et al. (2007, 2008) investigated several hundred simulated galaxy collisions and reported that star formation is indeed enhanced in interactions, but only by factors of a few, mainly occurring in nuclear starbursts, and driven primarily by large-scale inward gas flows driven by non-axisymmetries in the disks of the galaxies. 

Teyssier, Chapon \& Bournaud (2010), on the other hand, achieved much higher spatial resolution in their numerical modelling (a few tens of parsec, versus the approximately 200~pc in Di Matteo et al. 2007, 2008), which allowed them to resolved the cold and turbulent dust clouds. They deduced that the triggering mechanism of enhanced star formation in mergers is gas fragmentation into massive and dense clouds, rather than gas inflow. This led them to suggest that the SFRs induced in mergers can be up to a factor of ten higher than previously reported (e.g., by Di Matteo et al.) on the basis of numerical simulations with lower resolution.

Our observational confirmation that even though interacting galaxies show statistically enhanced SFRs as compared to a control population, their SFRs are only enhanced by a small factor (see also KJ09), lends support to the view that the higher level of star formation predicted by Teyssier et al. (2010) cannot be systematic, and may be limited to more extreme classes of merging galaxies such as, perhaps, ULIRGs. Alternatively, these models overpredict the observed SFRs by roughly an order of magnitude. 

\subsection{Galaxies with extreme relative star formation activity: starbursts?}

By their very nature of being extreme in terms of star formation, starburst galaxies are interesting. They produce large quantities of stars, of emission, and of stellar or starburst winds, all of which have important consequences for the cosmological and subsequent secular evolution of galaxies. Their detailed study is thus natural, and is helped by the fact that they are bright, and hence more easily observed than less prominently emitting galaxies, in particular at higher redshifts. 

Studies at low redshifts, such as the current one, can serve the important purpose of quantifying the relative importance of starbursts, in comparison to non-starburst galaxies, and across very wide ranges of galaxy mass and type. That starbursts occur across a range of galaxy properties is clear from Table~\ref{extremes}, which lists all galaxies with extreme SFR and SSFR values, as compared to their control populations. The Table indicates, firstly, that high values of SFR and SSFR go together (there are no galaxies with one of these parameters enhanced but not the other, although some values do not quite reach the cutoff value we used, of ten) and, secondly, that the galaxies identified here as starbursts span a wide range of type, though biased towards early types, and of mass, though excepting the lowest-mass galaxies, say below $\log M/M_{\odot}=9.5$. 

Another important conclusion from Table~\ref{extremes} is that, in particular in the early-type galaxies, the values of SFR and SSFR of many galaxies identified here as starbursts are particularly low in absolute terms. This was pointed out also by KJ09, and is due to the fact that the control values for early-type galaxies are very low. The implication is that a definition as used here for starbursts will lead to the inclusion of galaxies with relatively, but not necessarily absolutely, high SFRs and SSFRs, where those with low absolute values will have very limited impact on the evolution of the galaxies and their surroundings. The use of alternative starburst definitions, such as gas depletion time or absolute SFR, can, however, lead to other biases so are not necessarily any better (see KJ09 for a more detailed discussion). In the current context, where it is important to compare properties of interacting and non-interacting galaxies with control samples given the huge spread on galaxy properties in our sample, particularly mass and type, we consider the used definition as the most adequate, though noting our concerns.

\subsection{Interacting galaxies with low star formation}

An interesting and perhaps counterintuitive result that can be deduced from the top panels of Figure~\ref{enhan} is that a quarter of the most extreme interacting galaxies (Class A) have SFRs and SSFRs, normalised to their control samples, which are lower than unity. This means that those four galaxies have {\it SFRs and SSFRs which are below those in} the general population. In addition, another quarter is only marginally above unity in SFR and SSFR enhancement (within 1\,$\sigma$). Among Class B and C galaxies the situation is similar.

We can thus conclude that there exist significant sub-populations of interacting galaxies that have SFRs and SSFRs that are {\it lower than}, or not significantly {\it larger than, those found in} the control population. This also implies that the higher median SFR and SSFR enhancement among interacting galaxies, one of the main results of this paper, can never be interpreted in terms of a blanket statement such as `interacting galaxies have higher (S)SFRs'. Half of them do not, and only statistically, as a population, are interacting galaxies forming stars at higher rates. Timescales may be partly the issue here, as our observations do not show, e.g., whether an interaction has recently had, or will soon have, enhanced SFR/SSFR, or whether a galaxy with increased SFR/SSFR has lost its morphological signatures of being interacting. Our interaction classes likely contain a large spread in interaction stage and merger properties, and in future work we will explore this in more detail, as well as any relations with properties such as gas content or pair mass ratio.

\subsection{Studying nearby galaxies and implications for cosmological evolution}

Perhaps the most important rationale for carrying out the current study is using a statistically significant sample of local galaxies to investigate whether interactions affect the star formation properties of galaxies. With local we mean here within some 45\,Mpc distance, which is much nearer than studies in the literature which also refer to `local' galaxies, but include targets at many times our maximum distance. Using such nearby galaxies has two advantages: firstly, we can see relevant details such as tidal distortions in standard imaging, from SDSS and {\it Spitzer} in our case. Secondly, not only are fainter galaxies which dominate the general population in certain mass and type ranges included in the sample, we can also study companions to most of our sample galaxies which are some three magnitudes fainter. 

Studying such nearby galaxies leads to an immediate complication, though, which is that it becomes much harder to reach large samples. For instance, Luo et al. (2014) select galaxies with SDSS imaging at a redshift range from 0.01 to 0.20 and reach a sample size of almost 600,000 galaxies. In contrast, our sample is almost 1500, which seems small in comparison, but is the largest by far at the distance range of our galaxies ($\lesssim45$\,Mpc, $z<0.01$), and is large enough to draw important and statistically sound conclusions.

We thus provide a local baseline for studies of galaxies at slightly or much higher redshifts. As we discuss in more detail in Paper~III, quantities like the fraction of starburst galaxies and the absolute SFRs are lower in our study than in those of higher redshift galaxies, but this is only natural given the sample characteristics. As we confirm here, interacting galaxies indeed have higher SFRs, and are brighter, and they will thus be preferentially picked up in studies of more distant samples. It is thus vital to establish local values, as we are doing here.


\section{Conclusions}

We use a sample of almost 1500 nearby galaxies which form part of the extended S$^4$G survey and are closer than some 45\,Mpc to quantify the enhancement of the star formation by galaxy-galaxy interactions. We measure SFRs from {\it IRAS} fluxes. We use stellar masses determined from 3.6\,$\mu$m S$^4$G images corrected for emission from young stars and dust, and assuming a reliable $M/L$ ratio, confirmed by three independent methods, for the pure old stellar population left in the corrected images. Dividing the SFR by this stellar mass then yields the SSFR, which normalises the SFR by galaxy mass. We explore the distribution of the SFR and SSFR with morphological type and by galaxy stellar mass.

We calculate, for each galaxy, the SFR and SSFR normalised to the values obtained for a specific and individually crafted control sample, which is a specific population formed by all galaxies within a morphological type range of $\pm1$, and a stellar mass range of $\pm0.2$ in $\log(M/M_{\odot})$. We find that both SFR and SSFR are enhanced in interacting galaxies, and more so as the interaction becomes stronger. The increase is, however, moderate, of at most a factor of two. We discuss how those galaxies with the largest SFR and/or SSFR enhancement can be defined as starbursts, noting that although this definition selects a certain kind of starbursting galaxies, other starburst definitions may select different galaxies. We highlight that many interacting galaxies have SFRs and SSFRs that are not enhanced at all with respect to a control population, and in several cases even lower. We argue that this study based on a representative sample of nearby galaxies can be used to place strong constraints on studies based on samples of galaxies at larger distances. 


\section*{acknowledgments}

We thank Sara Ellison, Am\'elie Saintonge, Paola di Matteo, and Bruce Elmegreen for comments on an earlier version of the manuscript, and Alexandre Bouquin and Armando Gil de Paz for help with the UV SFRs. We also thank the entire S$^4$G team for their efforts in producing the S$^4$G data products. We acknowledge financial support to the DAGAL network from the People Programme (Marie Curie Actions) of the European Union's Seventh Framework Programme FP7/2007-2013/ under REA grant agreement number PITN-GA-2011-289313, and from the Spanish MINECO under grant number AYA2013-41243-P. This research made use of the NASA/IPAC Extragalactic Database which is operated by JPL, Caltech, under contract with NASA.  





\bsp

\label{lastpage}

\end{document}